\documentclass[namedreferences]{solarphysics}

\usepackage[hyperref,optionalrh]{spr-sola-addons} 
\usepackage{graphicx}        
\usepackage{amssymb}        
\usepackage{color}           

\chardef\us=`\_
\begin{document}
	
	\begin{article}
		\begin{opening}
			
			\title{Doppler Events in the Solar Photosphere:  The Coincident Superposition of Fast Granular Flows and p-mode Coherence Patches.\\ {\it Solar Physics}}
			
			\author[addressref={aff1,aff2},corref,email={rachel.mcclure@lasp.colorado.edu}]{\inits{R. L.}\fnm{R. Lee}~\lnm{McClure}}
			\author[addressref={aff1,aff2},corref]{\inits{M. P.}\fnm{Mark P.}~\lnm{Rast}}
			\author[addressref={aff3},corref]{\inits{V. M.}\fnm{Valentin }~\lnm{Mart{\'{\i}}nez Pillet}}
			
			\address[id=aff1]{Department of Astrophysical and Planetary Sciences, University of Colorado, Boulder CO 80309, USA}
			\address[id=aff2]{Laboratory for Atmospheric and Space Physics, University of Colorado, Boulder CO 80303, USA}
			\address[id=aff3]{National Solar Observatory, Boulder CO 80303, USA}
			
			\runningauthor{McClure et al.}
			\runningtitle{Doppler Events in the Solar Photosphere}
			
			\begin{abstract}
				Observations of the solar photosphere show spatially-compact large-amplitude Doppler velocity events with short lifetimes.  In data from the \textit{Imaging Magnetograph eXperiment} (IMaX) on the first flight of the \textsc{Sunrise} balloon in 2009, events with velocities in excess of 4$\sigma$ from the mean can be identified in both intergranular downflow lanes and granular upflows.  We show that the statistics of such events are consistent with the random superposition of strong convective flows and p-mode coherence patches. Such coincident superposition has implications for the interpretation of acoustic wave sources in the solar photosphere, and may be important to the interpretation of spectral line profiles formed in solar photosphere.
			\end{abstract}
			
			\keywords{
				Granulation;
				Oscillations, Solar;
				Waves, Acoustic
			}
		\end{opening}
		
		\section{Introduction}
		\label{sec:introduction} 
		
		Spectropolarimetric inversion of data from the \textit{Imaging Magnetograph eXperiment} (IMaX) instrument~\citep{pillet2010} on the first flight of the \textsc{Sunrise} stratospheric balloon~\citep{solanki2010} yielded two 20--30 minute high resolution time-series of the photospheric Doppler velocity with 33 second cadence.  These time-series show compact intermittent flashes of extreme Doppler values in the intergranular lanes and somewhat less conspicuous but complementary outstanding Doppler values within granules (see movie at \href{https://youtu.be/Z64rnq5yqW4}{\textit{link}}).
		
		Two physically distinct processes contribute to the magnitude of the Doppler velocity observed in the solar photosphere. Thermal convection is visible as granulation, with characteristic Doppler velocity amplitudes between $\pm0.5\,$--$1.5$ km s$^{-1}$ \citep{title1989}, and the random superposition of the solar acoustic oscillations \citep{ulrich1970} with maximum individual mode amplitudes of about $\pm 15$ cm s$^{-1}$ \citep[\textit{e.g.},][]{christensen2003} producing coherence patches with median vertical velocity amplitude of about $\pm0.5$ km s$^{-1}$ and maximum amplitudes of order $\pm1$ km s$^{-1}$ \citep{leighton1962,1988SSRv...47..275L}.  In this paper we study these two components of the flow at high spatial resolution, and show how together they yield the observed Doppler flashes.
		
		In Section 2 we describe the \textsc{Sunrise i} IMaX data in more detail and discuss the criteria used to identify the extreme Doppler velocity events.  In Section 3 we separate granular and p-mode contributions to the Doppler maps, examine the two components individually, and show that the number and amplitude distribution of extreme events are statistically consistent with the random superposition of locally strong granular flows and p-mode coherence patches.  Implications are discussed in Section 4.
		
		\section{Observations}
		\label{sec:data}
		
\begin{figure}[t!]
			\centering
			\includegraphics[width=\textwidth]{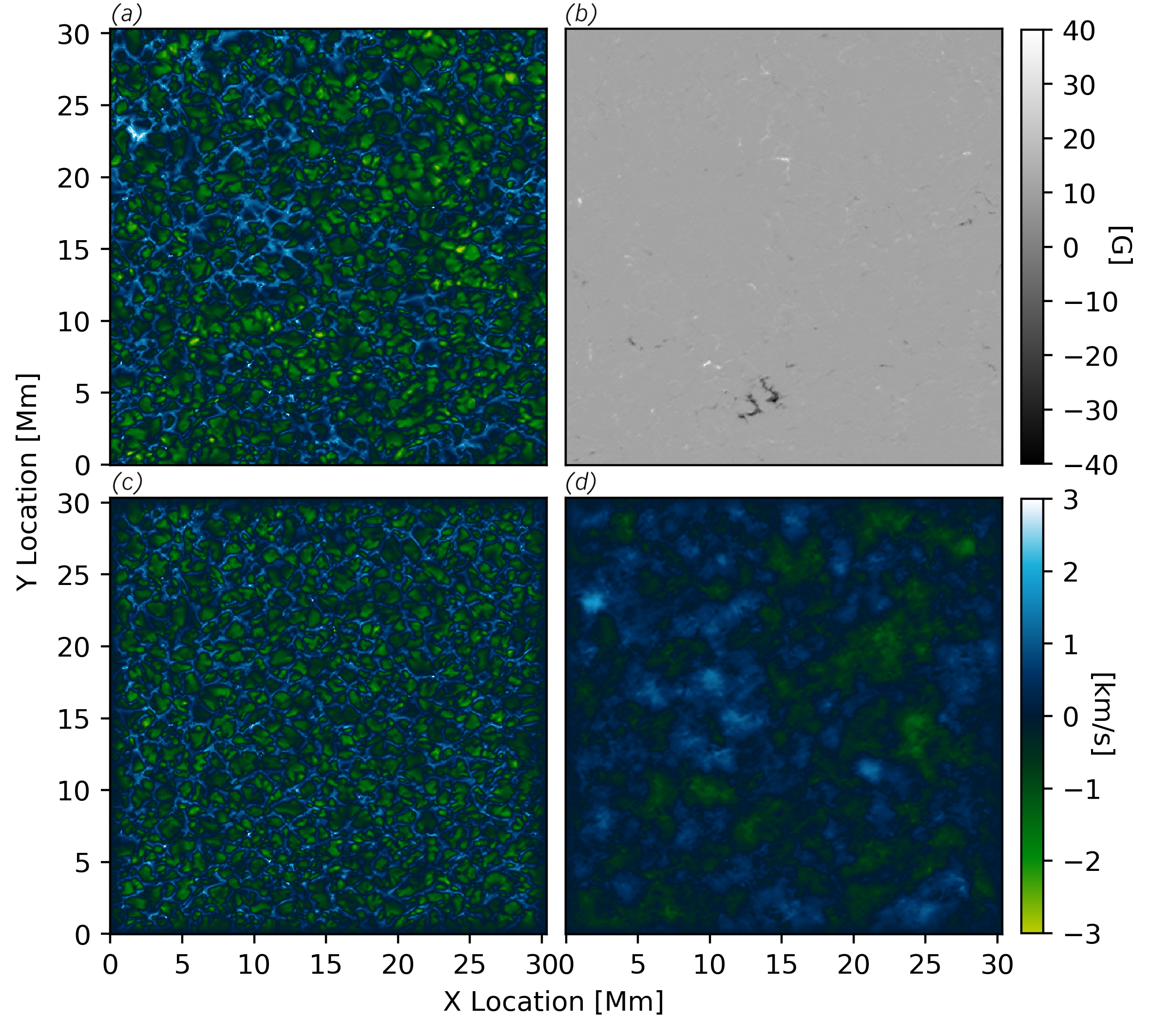}
			\caption{Single frame from the line of sight velocity time-series ($a$) derived from the IMaX spectrograph on the \textsc{Sunrise i} balloon flight.  Positive values correspond to down-flows and negative ones to up-flows. In ($c$) and ($d$), the granulation and p-mode contributions are shown using the same color scaling. Image ($b$) shows the longitudinal magnetic field over the same region with grey scale values ranging from -40 to 40 G.  The image is saturated to emphasize weak field.}
			\label{fig:og_gran_pmodes}
		\end{figure}

		The \textsc{Sunrise i} stratospheric balloon mission observed the Sun for 130 hours in 2009.  The balloon carried a 1 m Gregorian telescope with an effective focal length of $\sim25$ m at a float altitude of 35--40 km~\citep{solanki2010}.  The post-focus IMaX instrument used two custom nematic liquid crystal variable retarders, a commercial polarizing beamsplitter, and a double-pass LiNbO3 etalon~\citep{2006SPIE.6265E..2GA} to measured all four Stokes parameters within the Zeeman sensitive Fe \textsc{i} 525.02 nm line and in the neighboring continuum over a $50\times50$ arcsecond field of view with a cadence between 10 and 33 s~\citep{pillet2010}.  We present results from measurements taken at four spectral line positions (-80,-40,+40,+80 m\AA) and in the neighboring continuum (+227 m\AA).  Two regions near solar disk center were observed with spectral resolution of 85 m\AA\  at a cadence of 33 seconds, one for a total of 23 minutes and the other for 32 minutes (hereafter called image Set 1 and image Set 2 respectively).  A Milne-Eddington inversion (MILOS--~\citeauthor{2007PASJ...59S.837O}, \citeyear{2007PASJ...59S.837O}; \citeyear{suarez2010}) was used to determine the three magnetic field components, line of sight velocity, and plasma temperature over the field of view of each image frame in the two time-series.  Figure~\ref{fig:og_gran_pmodes}$a$ displays the line of sight velocity in a single frame.  Animations of the full time-series are available as supplementary material.
		
		The region under consideration is free of any large scale magnetism (see Figure \ref{fig:og_gran_pmodes}$b$).  Granulation is readily discernible, and acoustic oscillations are apparent as a time-dependent wavering across the field of view in the animated time-series and as large regions ($\sim2\times2$ Mm$^2$ patches) of enhanced or depressed vertical velocities in the still image of Figure~\ref{fig:og_gran_pmodes}$a$. 
		
		\subsection{Event Definition}
		\label{ssec:events}
		
		Spatially-compact large-amplitude Doppler velocity events of duration much shorter than the life-time of granules are observed at apparently random intervals and locations in the time-series.  These events occur with both signs; downflow events appear as striking flashes in the intergranular lanes and upflow events appear as high velocity localized sites within granules. With the color table selected for Figure \ref{fig:og_gran_pmodes}$a$, the events can be seen as compact bright white regions in the intergranular lanes (for example at $(x,y)=(2,24)$ in Figure \ref{fig:og_gran_pmodes}$a$) and bright yellow areas in the granules (for example at $(x,y)=(25,15)$ in Figure \ref{fig:og_gran_pmodes}$a$).  At the resolution of these observations the regions of strongest upflow are surprisingly compact.
		
		For analysis we define a Doppler event to be any region with an area of at least 10 pixels$^2$ ($\sim 400$ km$^2$) that shows velocities in excess of 4$\sigma$ above or below the temporal and spatial mean, over three or more consecutive frames.  This corresponds to velocities below -2.65 or above 1.98 km s$^{-1}$ in time-series Set 1 and below -2.71 or above 2.11 km s$^{-1}$ in Set 2.  The average event rate across the two time-series  is 1.9$\pm$0.3 upflow events and 2.0$\pm$0.3 downflow events per frame over the $30\times30$ Mm$^{2}$ central subregion retained after the apodization we employ in the Fourier filter described below.

		\subsection{Convective and Acoustic Contributions}
		\label{ssec:filter}
\begin{figure}
			\centering
			\includegraphics[width=\textwidth]{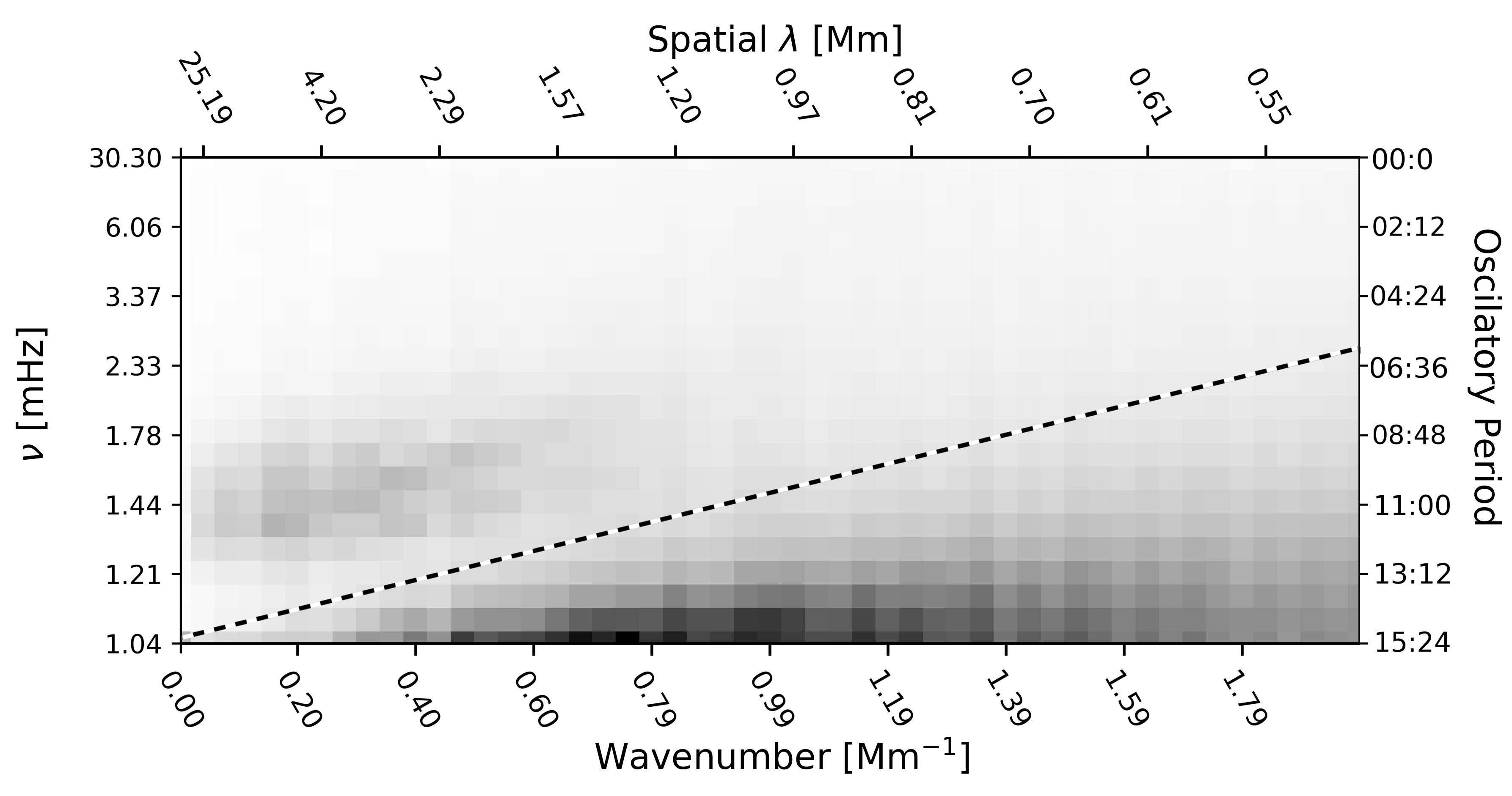}
			\caption{Average power spectrum of the two Doppler velocity time-series. The spectrum was azimuthally averaged in $k$, with annuli widths corresponding to the wavenumber resolution.  Application of a 5 km s$^{-1}$ filter (indicated by the {\it dashed line}), approximately separates granulation, below the line, from acoustic modes, above.}
			\label{fig:annulusPowerPlt}
		\end{figure}

		As discussed earlier, there are two contributions to the Doppler velocity measured at any position in the solar photosphere: the convective flow and the acoustic oscillations.  We separate these two contributions by applying a Fourier filter to the Doppler data ({\it e.g.,} \citeauthor{hill1988}, \citeyear{hill1988}; \citeauthor{schou1998}, \citeyear{schou1998}).  The images were spatially apodized using a two-dimensional cosine-bell taper of width 39 pixels applied to each edge, Fourier transformed in time and space, and separated into the granular and p-mode components either side of 5 km s$^{-1}$ (see Figure~\ref{fig:annulusPowerPlt}).  Single frames of the resulting granular and acoustic contributions are shown in Figure~\ref{fig:og_gran_pmodes}$c$ and \ref{fig:og_gran_pmodes}$d$. 
		
\begin{figure}
			\centering
			\includegraphics[width=.95\textwidth]{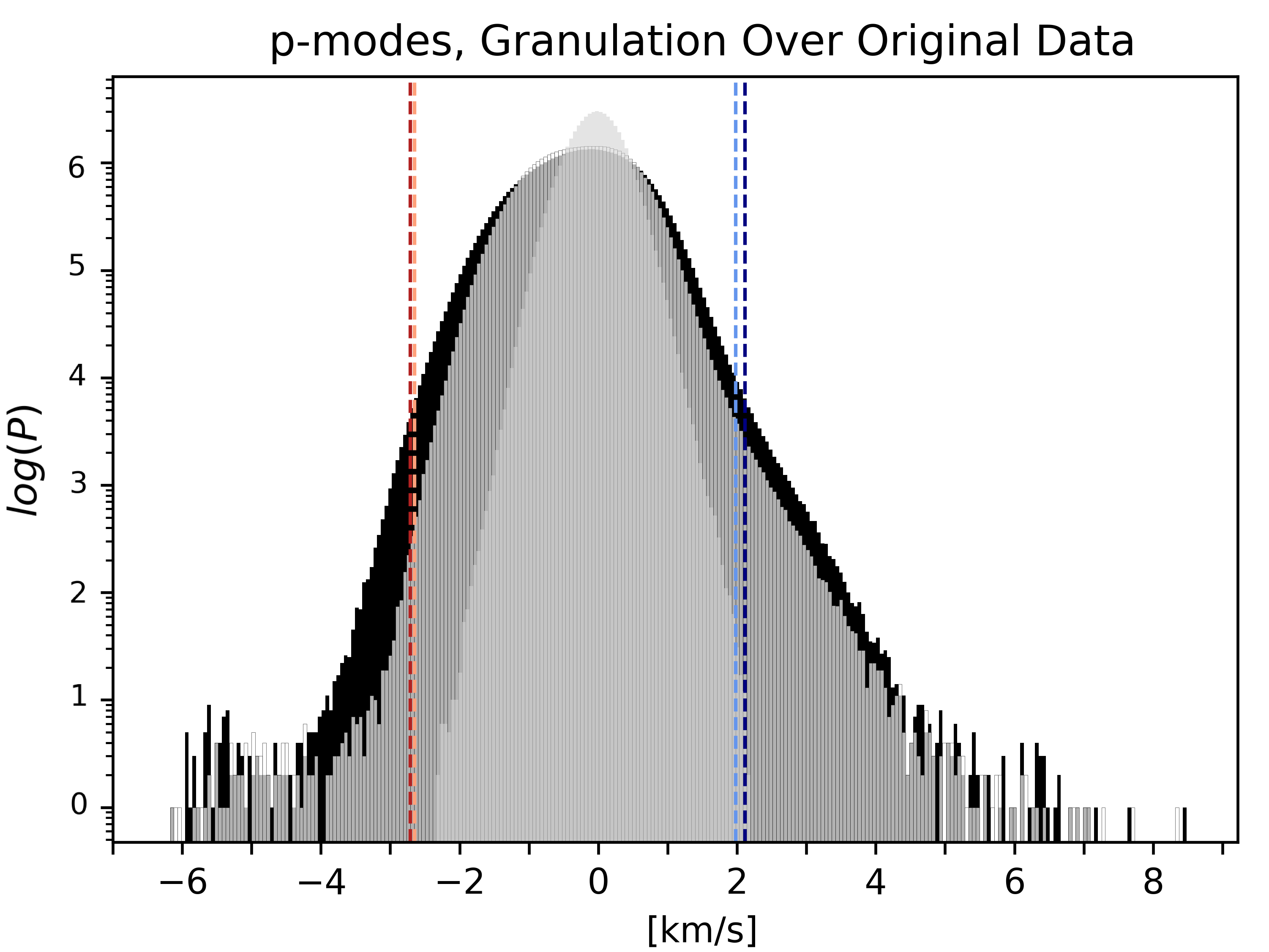}
			\caption{Doppler velocity distributions, with the  p-mode and granulation contributions plotted in {\it light} and {\it dark grey} respectively, and the total distribution shown in {\it black}. Dashed lines are $\pm 4\sigma$ from the mean of each observed set (Set 1 upflow is orange, downflow is light blue and Set 2 upflow is red, downflow is dark blue). The p-mode contribution is normally distributed about zero, while the granulation shows enhanced high speed tails of both signs.}
			\label{fig:dataPG}
		\end{figure}

		The distributions of Doppler velocities measured for each component are shown in Figure \ref{fig:dataPG}.  Not surprisingly, the p-mode contribution is normally distributed about zero, since it results from the random superposition of oscillatory symmetric eigenfunctions.  On the other hand, the distribution of the convective velocities is significantly non-Gaussian, with enhanced occurrence of large amplitude flows of both signs.  The mean is shifted to negative values because the area of the upflowing regions exceeds that of the downflows, and the distribution is asymmetric, with extreme values more prevalent in downflowing regions (positive values in Figure \ref{fig:dataPG}).  The median upflow speed (0.60 km s$^{-1}$ for Set 1 and 0.61 km s$^{-1}$ for Set 2) is higher than the median downflow speed (0.43 km s$^{-1}$ for Set 1 and 0.45 km s$^{-1}$ for Set 2), and notably, at the resolution of these observations, granular upflows are highly structured with peak upflow speeds in small regions approaching those found in downflow lanes.   
		
		\section{Coincident Superposition}
		\label{sec:testing}
		To investigate the origin of the extreme Doppler velocity events identified, we test the hypothesis that such events are caused by the random superposition of the p-mode and granular contributions. We create synthetic data sets by superimposing the observed p-mode and granulation fields, spatially offsetting the two components and co-adding them to form synthetic image time-series. Each frame capturing the p-mode contribution is periodically shifted in increments of 20 pixels 32 times, separately horizontally and vertically and simultaneously in both directions. This process creates 102 synthetic observations at each time step.  Since each frame is over 700 pixels on a side and the minimum event size is taken to be 10 pixels$^2$, shifts of 20 pixels avoids double counting synthetic events.  
		
		\begin{figure}[h!]
			\centering
			\includegraphics[width=.9\textwidth]{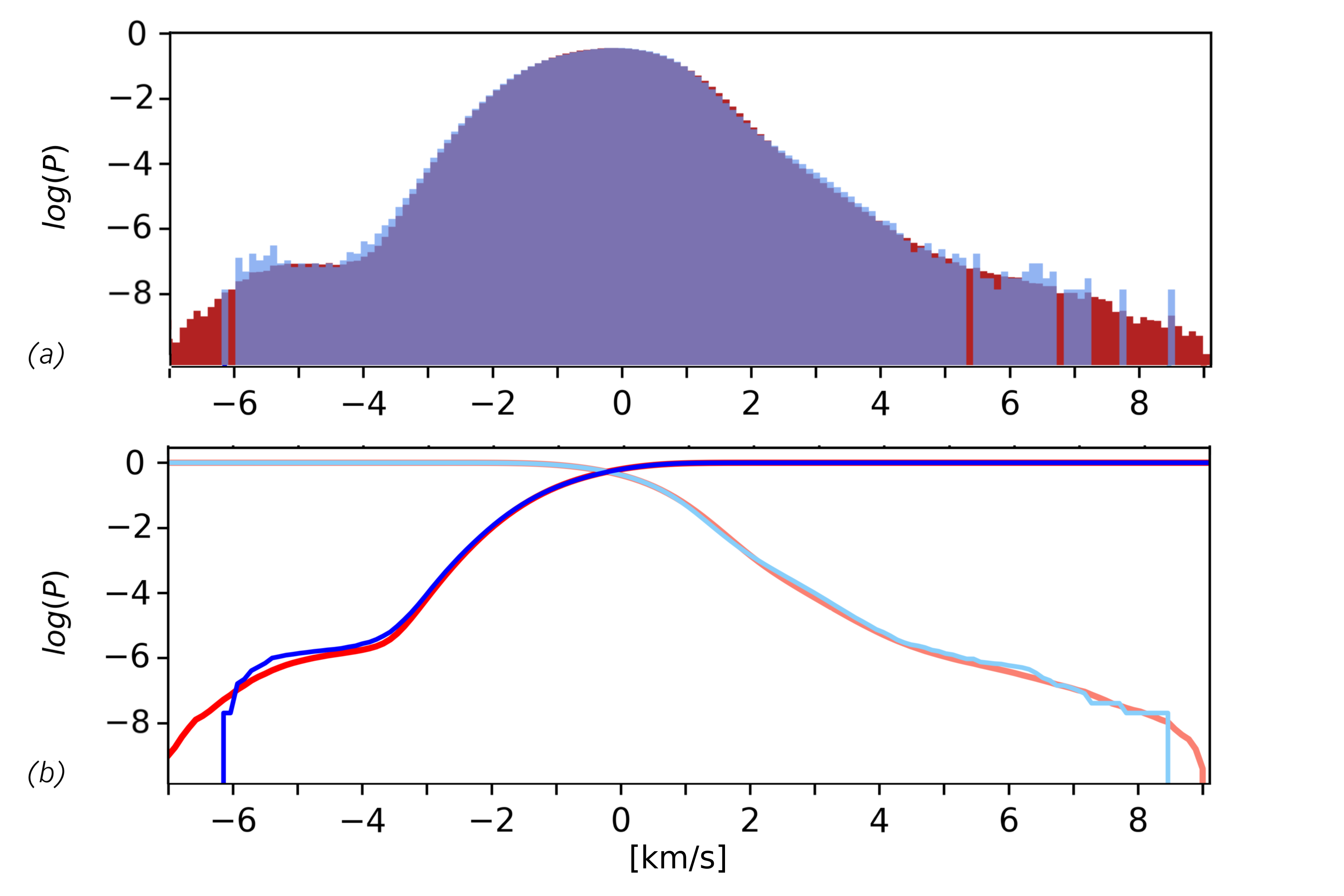}
			\caption{The normalized probability density of Doppler amplitudes ($a$) in the synthetic data sets produced by random superposition of the p-mode and convective contributions ({\it red}) and the observed time-series ({\it blue}). Normalized cumulative distributions ($b$) of the Doppler velocities, taken from the upflow (negative value) side, left-to-right, and from the downflow (positive value) side, right-to-left, of the observed data in {\it blue} and model data in {\it red}.}
			\label{fig:cdfAndHist}
		\end{figure}

		\subsection{Event Amplitude Distributions}
		\label{ssec:model}
		
		The model and the original data Doppler amplitude distributions are quite similar (Figure \ref{fig:cdfAndHist}), with some small differences apparent in the distribution shoulders and more noise in the observed distribution tails than in the synthetic data for which more time-series are available.  What is however more constraining than a comparison between the velocity distributions is an assessment of event rates, since this captures spatial and temporal coherences within the Doppler map via the event definition. 
		
\begin{figure}[h]
			\includegraphics[width=\textwidth]{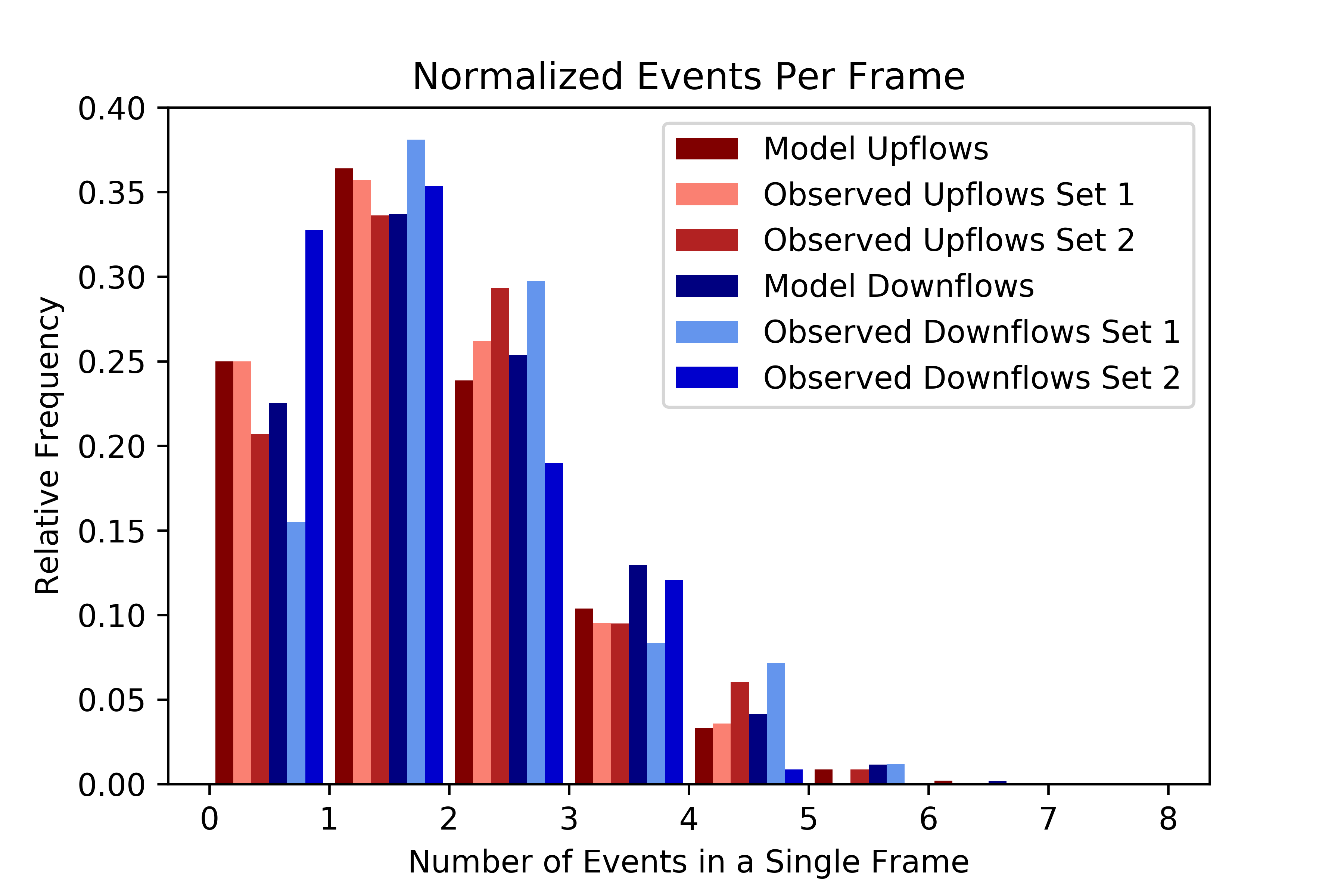}
			\caption{Probability density of the number of event counts per frame in the observations and the random superposition model.}
			\label{fig:histall}
		\end{figure}

		The overall mean event rates are listed in the Table \ref{table:counts} of the appendix. The observed mean event rate lies within 2$\sigma$ of the mean rate in the synthetic time-series produced by the random superposition of granular and p-mode motions.  The probability densities of the number of counts per frame (Figure~\ref{fig:histall}) also agrees quite well with that of the observed data, but because of the low number of total counts in the observed data, it is difficult to tell whether the differences are significant.

		\subsection{Occurrence Frequency}
		\label{ssec:stats}
		
		\begin{figure}[h!]
			\includegraphics[width=\textwidth]{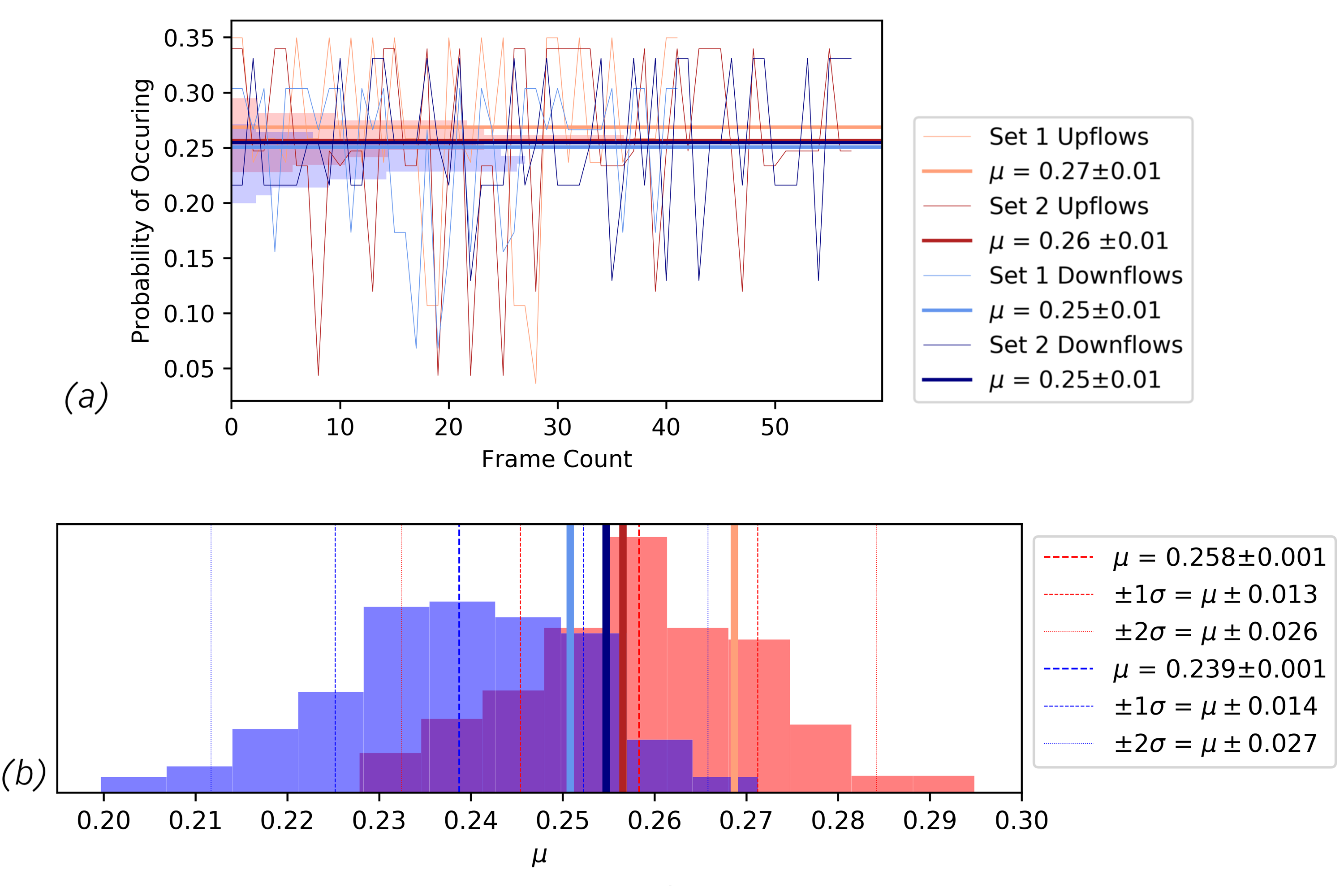}
			\caption{Poisson probability of observing the number of events found in each Doppler velocity image. In $a$, that probability as a function of frame number for the observed time-series, and the distribution of those probabilities for upflow and downflow events in both simulations projected along the y-axis.  In $b$, the distribution of the Poisson probability in upflows and downflows of the a synthetic model data.  The mean Poisson probability values in the observations are indicated with solid fiducial lines in both panels.  The mean, $\pm1\sigma$ and $\pm2\sigma$ Poisson probability values of the model distributions are indicated with {\it vertical dashed and dotted lines} in ($b$).}
			\label{fig:poissonCts}
		\end{figure}
		
		To examine this further we look at the probability of the number of events observed in each frame.  Since event counts are small the distribution approximates a Poisson distribution, so we calculate for each each frame the Poisson probability of finding the number of events observed if the mean is that determined from the random superposition model.  Figure \ref{fig:poissonCts}$a$ plots that probability as a function of frame number for the observed time-series, along with the distributions of those values for upflow and downflow events in both simulations projected along the y-axis.  Mean Poisson probability values are indicated with solid horizontal fiducial lines.  Similarly in Figure \ref{fig:poissonCts}$b$, we plot the distribution of the Poisson probability of finding the specific number of events in each frame of a synthetic data set given the mean number of event counts from all the other model sets.  In this way, each model time-series is treated in the same manner as the observational ones.  
		
		The mean values of the Poisson probabilities of the observed counts (indicated in Figure \ref{fig:poissonCts}$b$ by solid vertical fiducial lines) falls at or within 1$\sigma$ of the means of the synthetic data Poisson probability distributions. This is true of all superposition models, both upflows and downflows in all observation time-series. We conclude that the number of outstanding Doppler events in each frame of the observational data is statistically consistent with counts that would be achieved by the random superposition of p-mode and the granular flows.

		\section{Conclusion}
		\label{sec:conclusion}
		The random superposition of solar acoustic oscillation coherence and underlying convective flows can cause intermittent Doppler signals that have amplitudes in excess of 4$\sigma$ from the mean, values that are thus faster than 99.7\% of those observed if the distribution is approximated as Gaussian. These Doppler events occur in both up-flowing and down-flowing regions, and flows of this magnitude are three-times more likely to occur in an unfiltered time-series than in one from which the p-modes have been removed. It is important to note that these events are real physical occurrences, not an artifact of the observations. Their existence may be critical to the interpretation of spectral lines, the identification of solar acoustic sources, and the understanding of jet-like structures. They may cause confusion in interpreting observations as instances of specific dynamical mechanisms, particularly when observational time-series are too short to adequately remove the p-mode contributions. 
		
		\begin{acks}
			This work was supported in part by the National Science Foundation under grant No. 1616538. The National Solar Observatory is operated by the Association of Universities for Research in Astronomy under a cooperative agreement with the National Science Foundation. Disclosure of Potential Conflicts of Interest: The authors declare that they have no conflicts of interest.
			
		\end{acks}
		
		\newpage
		\textbf{Appendix}
		
		Here we tabulate the observed and synthetic event counts, showing that the event statistics are consistent with the random superposition of acoustic mode coherence patches and strong granular flows. Comparable events rates are found in upflows and downflows. This is a consequence of being able to resolve the upflow substructure at \textsc{Sunrise} resolutions.
		\begin{table}[h!]
			\caption{Event counts and event rates for observed data, granular and p-mode components, and the superposition model sets.}
			\begin{tabular}{ |p{3.3cm}p{.7cm}||p{2.4cm}||p{2.4cm}|}
				\hline
				\textbf{Data Component}&	&Number of Events&Mean Event Rate\\
				&	&			&s$^{-1}$ Mm$^{-2}$ \\
				\hline
				\hline
				\textbf{Observed Data}&&&\\
				Upflows 	&Set 1		&	74	& 5.6$\times10^{-5}\pm$9\%	\\
				&Set 2		&	123	& 6.8$\times10^{-5}\pm$7\%	\\
				Downflows 	&Set 1		&	104	& 5.0$\times10^{-5}\pm$10\%	\\
				&Set 2		&	99	& 3.6$\times10^{-5}\pm$10\%	\\
				\hline
				\textbf{Apodized Data}&&&\\
				Upflows 	&Set 1		&	55	& 5.1$\times10^{-5}\pm$10\%\\
				&Set 2		&	86	& 5.8$\times10^{-5}\pm$10\%\\
				Downflows 	&Set 1		&	66	& 6.2$\times10^{-5}\pm$10\%\\
				&Set 2		&	65	& 4.4$\times10^{-5}\pm$10\%\\
				\hline
				\textbf{Granular Component}&&&\\
				Upflows 	&Set 1		&	 4	& 3.7$\times10^{-6}\pm$60\%\\
				&Set 2		&	 3	& 2.4$\times10^{-6}\pm$60\%\\
				Downflows 	&Set 1		&	22	& 2.1$\times10^{-5}\pm$20\%\\
				&Set 2		&	14	& 9.8$\times10^{-6}\pm$30\%\\
				\hline
				\textbf{p-mode Component}&&&\\
				Upflows 	&Set 1	 	& 	0	& 0				\\%
				&Set 2	 	& 	0	& 0				\\%
				Downflows	&Set 1	 	& 	0	& 0				\\%
				&Set 2	 	& 	0	& 0				\\%
				\hline
				\textbf{Model}&&&\\
				Upflows 	&Set 1	 	& 1.6$\pm$.1/frame	& 5.30$\times10^{-5}\pm$1\%	\\%
				&Set 2	 	& 1.7$\pm$.1/frame	& 5.70$\times10^{-5}\pm$1\%	\\%
				Downflows 	&Set 1		& 2.0$\pm$.2/frame	& 6.87$\times10^{-5}\pm$1\%	\\%
				&Set 2		& 1.8$\pm$.1/frame	& 6.00$\times10^{-5}\pm$1\%	\\%
				\hline
			\end{tabular}
			
			\label{table:counts}
		\end{table}
		
		\section*{}
		\newpage
		\bibliographystyle{spr-mp-sola}
		\bibliography{refs}	
		
	\end{article} 
	
\end{document}